\documentstyle[12pt]{article}

\begin{document}
{\sf \begin{center} \noindent
{\Large \bf Lagrangean stable slow dynamos in compact 3D Riemannian manifolds}\\[3mm]

by \\[0.3cm]

{\sl L.C. Garcia de Andrade}\\

\vspace{0.5cm} Departamento de F\'{\i}sica
Te\'orica -- IF -- Universidade do Estado do Rio de Janeiro-UERJ\\[-3mm]
Rua S\~ao Francisco Xavier, 524\\[-3mm]
Cep 20550-003, Maracan\~a, Rio de Janeiro, RJ, Brasil\\[-3mm]
Electronic mail address: garcia@dft.if.uerj.br\\[-3mm]
\vspace{2cm} {\bf Abstract}
\end{center}
\paragraph*{}
Modifications on a recently introduced fast dynamo operator by
Chiconne et al [Comm Math Phys 173, 379 (1995)] in compact 3D
Riemannian manifolds allows us to shown that slow dynamos are
Lagrangean stable, in the sense that the sectional curvature of the
Riemann manifold vanishes. The stability of the holonomic filament
in this manifold will depend upon the sign of the second derivative
of the pressure along the filament and in the non-holonomic case, to
the normal pressure of the filament. Lagrangean instability is also
investigated in this case and again an dynamo operator can be
defined in this case. Negative curvature (Anosov flows) dynamos are
also discussed in their stability aspects. \vspace{0.5cm} \noindent
{\bf PACS numbers:} \hfill\parbox[t]{13.5cm}{02.40}

\newpage
\section{Introduction}
In chaotic dynamos \cite{1,2}, Lyapunov exponents \cite{3} or Arnold
\cite{2} Zeldovich \cite{4} stretching are strongly responsible for
the dynamo instability in an infinitely conducting fluid flow, as
discussed earlier by Friedlander and Vishik \cite{5}. This fluid
possesses a magnetic Reynolds number $R_{m}\rightarrow\infty$ which
is inversely proportional to the resistivity number ${\eta}$.
Therefore to better understand the process of instabilities is
important to investigate the relation between these dynamo operators
and the instabilities in the dynamo flows. Since , as pointed out by
Kambe \cite{6} the Lagrangean instability of flows is one of the
most important geometrical and topological instabilities in fluid
dynamics, a throughly understanding of this relationship is certain
important for dynamo theory. With this motivation , in this paper we
define a new dynamo operator in 3D compact Riemannian manifold which
as is shown in the form of the theorem $1$ below, which help us to
show that slow dynamos are stable in the Lagrangean sense \cite{6}.
Chiconne et al \cite{7} have recently obtained a similar operator
and investigated its fast dynamo spectrum and the corresponding
spectrum of group acting in the space of continuos divergence free
vector field $\vec{u}$ and $\vec{H}$ representing, respectively the
flow and the magnetic field in this compact Riemann manifold. Here
the Lagrangean stability in this very same space is computed and it
is shown that the spectrum eigenvalue equation in the form of the
self-induction equation, leads to stable slow dynamos. Slow dynamos
are defined with constraint $(p_{0}=0)$ , where the magnetic field
is given by the exponential stretching
$\vec{H}={\vec{H}}_{0}e^{p_{0}t}$ is given. As a corolary it is
shown that in the case of magnetic twisted filaments the stability
of the dynamos depend upon the second derivatives of the flow
pressure along the filaments. From the physical viewpoint there are
working examples of slow dynamos which are much easier to obtain
that the kinematic fast dynamos. This discussion was first given by
Soward and Childress \cite{8} and Moffatt and Proctor \cite{9}. The
paper is organized as follows: In section 2 a review of Riemann
sectional curvature is given. In section 3 dynamo operator spectrum
is computed. In this same section magnetic filament examples in
holonomic and nonholonomic cases are examined concerning its
Lagrangean stability properties. Section 4 we present the
conclusions.
\section{Sectional Riemann curvature}
 In this section we make a brief review of the differential geometry of surfaces in coordinate-free language of differential geometry.
 The Riemann curvature is defined by
 \begin{equation}
 R(X,Y)Z:={\nabla}_{X}{\nabla}_{Y}Z-{\nabla}_{Y}{\nabla}_{X}Z-{\nabla}_{[X,Y]}Z\label{1}
 \end{equation}
where $X {\epsilon} T\cal{M}$ is the vector representation which is
defined on the tangent space $T\cal{M}$ to the 3D Riemannian
manifold $\cal{M}$. Here ${\nabla}_{X}Y$ represents the covariant
derivative given by
\begin{equation}
{\nabla}_{X}{Y}= (X.{\nabla})Y\label{2}
 \end{equation}
which for the physicists is intuitive, since we are saying that we
are performing derivative along the X direction. The expression
$[X,Y]$ represents the commutator, which on a vector basis frame
${\vec{e}}_{l}$ in this tangent sub-manifold defined by
\begin{equation}
X= X_{k}{\vec{e}}_{k}\label{3}
\end{equation}
or in the dual basis ${{\partial}_{k}}$
\begin{equation}
X= X^{k}{\partial}_{k}\label{4}
\end{equation}
can be expressed as
\begin{equation}
[X,Y]= (X,Y)^{k}{\partial}_{k}\label{5}
\end{equation}
In this same coordinate basis now we are able to write the curvature
expression (\ref{1}) as
\begin{equation}
R(X,Y)Z:=[{R^{l}}_{jkp}Z^{j}X^{k}Y^{p}]{\partial}_{l}\label{6}
\end{equation}
where the Einstein summation convention of tensor calculus is used.
The expression $R(X,Y)Y$ which we shall compute bellow is called
Ricci curvature. The sectional curvature which is very useful in
future computations is defined by
\begin{equation}
K(X,Y):=\frac{<R(X,Y)Y,X>}{S(X,Y)}\label{7}
\end{equation}
where $S(X,Y)$ is defined by
\begin{equation}
{S(X,Y)}:= ||X||^{2}||Y||^{2}-<X,Y>^{2}\label{8}
\end{equation}
where the symbol $<,>$ implies internal product.
 \section{Dynamo operator for twisted filaments}
 Let us now start by considering the MHD field equations
and the framework for non-holonomic filaments \cite{10}
\begin{equation}
{\nabla}.\vec{H}=0 \label{9}
\end{equation}
\begin{equation}
{\nabla}{\times}{\vec{H}}= {{\partial}_{t}}\vec{H} \label{10}
\end{equation}
the field $\vec{H}$ along the filament. The vectors $\vec{t}$ and
$\vec{n}$ along with binormal vector $\vec{b}$ together form the
Frenet frame which obeys the Frenet-Serret equations
\begin{equation}
\vec{t}'=\kappa\vec{n} \label{11}
\end{equation}
\begin{equation}
\vec{n}'=-\kappa\vec{t}+ {\tau}\vec{b} \label{12}
\end{equation}
\begin{equation}
\vec{b}'=-{\tau}\vec{n} \label{13}
\end{equation}
the dash represents the ordinary derivation with respect to
coordinate s, and $\kappa(s,t)$ is the curvature of the curve where
$\kappa=R^{-1}$. Here ${\tau}$ represents the Frenet torsion. We
follow the assumption that the Frenet frame may depend on other
degrees of freedom such as that the gradient operator becomes
\begin{equation}
{\nabla}=\vec{t}\frac{\partial}{{\partial}s}+\vec{n}\frac{\partial}{{\partial}n}+\vec{b}\frac{\partial}{{\partial}b}
\label{14}
\end{equation}
 The other equations for the other legs of the Frenet frame are
\begin{equation}
\frac{\partial}{{\partial}n}\vec{t}={\theta}_{ns}\vec{n}+[{\Omega}_{b}+{\tau}]\vec{b}
\label{15}
\end{equation}
\begin{equation}
\frac{\partial}{{\partial}n}\vec{n}=-{\theta}_{ns}\vec{t}-
(div\vec{b})\vec{b} \label{16}
\end{equation}
\begin{equation}
\frac{\partial}{{\partial}n}\vec{b}=
-[{\Omega}_{b}+{\tau}]\vec{t}-(div{\vec{b}})\vec{n}\label{17}
\end{equation}
\begin{equation}
\frac{\partial}{{\partial}b}\vec{t}={\theta}_{bs}\vec{b}-[{\Omega}_{n}+{\tau}]\vec{n}
\label{18}
\end{equation}
\begin{equation}
\frac{\partial}{{\partial}b}\vec{n}=[{\Omega}_{n}+{\tau}]\vec{t}-
\kappa+(div\vec{n})\vec{b} \label{19}
\end{equation}
\begin{equation}
\frac{\partial}{{\partial}b}\vec{b}=
-{\theta}_{bs}\vec{t}-[\kappa+(div{\vec{n}})]\vec{n}\label{20}
\end{equation}
Let us now consider the main result of the paper in the form of a

\textbf{Theorem 1}: Let $\cal{M}$ be a compact Riemannian 3D
manifold and {L} the dynamo operator defined by:

\begin{equation}
{L}: \vec{u}\rightarrow (\vec{u}.{\nabla}) \label{21}
\end{equation}
Thus the slow dynamo condition $p_{0}=0$ yields the Lagrangean
stability given by the vanishing of the sectional curvature
$K(\vec{u},\vec{H})$. Fast dynamos can also be stable if the
condition $S>0$ is fulfilled. Here the proof shall be done with the
help of the following eigenvalue spectrum equation
\begin{equation}
L\vec{H}=p_{0}\vec{H} \label{22}
\end{equation}
$\vec{H}$ obeys the self-induction equation and
$R_{m}\rightarrow{\infty}$ for a highly conducting fluid.

\textbf{Proof}: By writing the self-induction equation in the form
\cite{8}
\begin{equation}
{\partial}_{t}\vec{H}={R^{-1}}_{m}{\nabla}^{2}\vec{H}+[\vec{H},\vec{u}]
\label{23}
\end{equation}
where $[\vec{H},\vec{u}]$ is the bracket defined in Riemannian 3D
manifold as
\begin{equation}
[\vec{H},\vec{u}]=
{\nabla}_{\vec{H}}\vec{u}-{\nabla}_{\vec{u}}\vec{H}\label{24}
\end{equation}
Taking into account the expression for the Riemann curvature in
section II allow us to express the dynamo equation in terms of the
Riemann curvature as
\begin{equation}
({\partial}_{t}\vec{H}).{\nabla}\vec{u}
=R(\vec{H},\vec{u})\vec{u}-{\nabla}_{\vec{H}}{\nabla}_{\vec{u}}\vec{u}+{\nabla}_{\vec{u}}{\nabla}_{\vec{H}}\vec{u}+
{R^{-1}}_{m}{{\Delta}}{\vec{H}}.{\nabla}\vec{u}\label{25}
\end{equation}
where ${\Delta}:={\nabla}^{2}$. From these expressions one is able
to compute the sectional curvature as
\begin{equation}
(\vec{H}.{\nabla}){\nabla}p-R^{-1}(\vec{H}.{\nabla}){\Delta}\vec{u}-
{R^{-1}}_{m}({{\Delta}}{\vec{H}}.{\nabla})\vec{u}=R(\vec{H},\vec{u})\vec{u}\label{26}
\end{equation}
where we have used the Navier-Stokes viscous flow equation
\begin{equation}
-{\nabla}p+R^{-1}{\Delta}{\vec{u}}={\partial}_{t}\vec{u}\label{27}
\end{equation}
and $R$ is the flow Reynolds number. Sectional curvature is then
\begin{equation}
K(\vec{H},\vec{u})=\frac{<(\vec{H}.{\nabla}){\nabla}p,\vec{H}>}{S(\vec{H},\vec{u})}=\frac{<R(\vec{H},\vec{u})\vec{u},\vec{H}>}{S(\vec{H},\vec{u})}\label{28}
\end{equation}
\begin{equation}
K(\vec{H},\vec{u})=\frac{<L{\nabla}p,\vec{H}>}{S(\vec{H},\vec{u})}=p_{0}\frac{||\vec{H}||^{2}}{S}\label{29}
\end{equation}
From this last expression the theorem is proved, since $S$ does not
vanish which implies $p_{0}=0$. The norm $||...||$ is built from the
Riemannian compact metric in the 3D manifold. Note that our
computations are also in agreement with Kambe assertion that the
Riemann connection would not be flat when pressure does not vanish.

Let us now consider the holonomic frame described in previous
section to write the magnetic filament in the form
$\vec{H}=H_{s}(s,t)\vec{t}$, which leads to the following result
\begin{equation}
R(\vec{H},\vec{u})={<(\vec{H}.{\nabla}){\nabla}p,\vec{H}>}={H_{s}}^{2}{{\partial}_{s}}^{2}p\label{30}
\end{equation}
where the use has been made of grad operator in the form
\begin{equation}
{\nabla}=\vec{t}{\partial}_{s}\label{31}
\end{equation}
Thus from expression (\ref{31}) one notes that the dynamo could be
unstable if the second derivative of the filament pressure is
positive. In the non-holonomic case is more complicated cause the
Riemann tensor could be negative , since
\begin{equation}
<R(\vec{H},\vec{u})\vec{u},\vec{H}>={H_{s}}^{2}[{{\partial}_{s}}^{2}p-{\kappa}(s){\partial}_{n}p]\label{32}
\end{equation}
where now the grad operator used in the last computation
was
\begin{equation}
{\nabla}=\vec{t}{\partial}_{s}+\vec{n}{\partial}_{n}+\vec{b}{\partial}_{b}\label{33}
\end{equation}
here in cases in which the Frenet curvature is strong as in some
plasma loops in the sun \cite{11}, for example, the RHS of this
equation can be negative and the Lagrangean instability of the
dynamo would carachterize an Anosov flow.
\section{Conclusions}
 In conclusion, we show that a dynamo operator defined on a compact 3D Riemannian manifold  lead us to show that the slow dynamos
 are Lagrangean stable. It is noted that fast dynamo filaments could also be stable when the pressure of the filaments is appropriatly
 constrained. Twisted filaments in non-holonomic frame may lead to
 filamentary dynamos which can be unstable when the Frenet curvature
 of the plasma solar loops , for example, is too strong. A simple
 generalization of the work here can be made by generalizing dynamos
 to non-Riemannian manifolds where Cartan torsion is considered as
 an obstruction of the self-induction equation.

 \section*{Acknowledgements}
 Thanks are due to CNPq and UERJ for financial supports.

\end{document}